# Relevance distributions across Bradford Zones: Can Bradfordizing improve search?

Philipp Mayr[1]

[1] philipp.mayr@gesis.org
GESIS – Leibniz Institute for the Social Sciences, Unter Sachsenhausen 6-8, 50667 Cologne, Germany

**Abstract**
The purpose of this paper is to describe the evaluation of the effectiveness of the bibliometric technique Bradfordizing in an information retrieval (IR) scenario. Bradfordizing is used to re-rank topical document sets from conventional abstracting & indexing (A&I) databases into core and more peripheral document zones. Bradfordized lists of journal articles and monographs will be tested in a controlled scenario consisting of different A&I databases from social and political sciences, economics, psychology and medical science, 164 standardized IR topics and intellectual assessments of the listed documents. Does Bradfordizing improve the ratio of relevant documents in the first third (core) compared to the second and last third (zone 2 and zone 3, respectively)? The IR tests show that relevance distributions after re-ranking improve at a significant level if documents in the core are compared with documents in the succeeding zones. After Bradfordizing of document pools, the core has a significant better average precision than zone 2, zone 3 and baseline. This paper should be seen as an argument in favour of alternative non-textual (bibliometric) re-ranking methods which can be simply applied in text-based retrieval systems and in particular in A&I databases.

**Introduction**
The perceived expectations of users searching the web are that retrieval systems should list the most relevant or valuable documents in the result list first (so-called relevance ranking). More approaches appear that draw on advanced methods to produce relevant results and alternative views on document spaces. Google PageRank and its derivations (see e.g. Lin, 2008) or Google Scholar's citation count are just two popular examples for informetric-based rankings applied in Internet search engines.

Distributed search across multiple A&I databases will also generate large and heterogeneous document sets with the effect that users are confronted with a massive load of results from different scientific domains, even for specific research topics. Furthermore, empirical tests with typical A&I databases like Medline show that conventional term frequency - inverse document frequency (tf-idf) best match models and especially recent web-based ranking methods implemented in search engines (originally for web pages) are not always appropriate for search in heterogeneously collected scholarly metadata documents.

In this paper we want to apply and evaluate a non-textual ranking technique, called Bradfordizing. Introduced by H.D. White (1981), Bradfordizing is a bibliometric method to reorganize a search result for a topic. Bradfordizing is set up by applying the following procedure:

"… that is sorting hits (1) by the journal in which they appear, and then sorting these journals not alphabetically by title but (2) numerically, high to low, by number of hits each journal contains. In effect, this two-step sorting ranks the search output in the classic Bradford manner, so that the most productive, in terms of its yield of hits, is placed first; the second-most productive journal is second; and so on, down through the last rank of journals yielding only one hit apiece." (White, 1981: p. 47).

*Bradford Law*
Journals play an important role in the scientific communication process. They appear periodically, they are topically focused, they have established standards of quality control and often they are involved in the academic gratification system. Metrics like the famous impact factor are aggregated on the journal level. In some disciplines journals are the main place for a scientific community to communicate and discuss new research results. These examples shall illustrate the impact journals bear in the context of science models (Börner et al., 2011). Modeling science or understanding the functioning of science has a lot to do with journals and journal publication characteristics. These journal publication characteristics are the point where Bradford law can contribute to the larger topic of science models.

Bradford law of scattering bases on literature observations the librarian S. Bradford has been carried out in 1934. His findings and after that the formulation of the bibliometric model stand for the beginning of the modern documentation (Bradford, 1948) – a documentation which founds decisions on quantifiable measures and empirical analyses. The early empirical laws described by Lotka, Zipf and of course Bradford are landmark publications which still influence research in scientometrics (Bookstein, 1990), but also in other research communities like computer science or linguistics. In brief, scientometric and informetric research investigates the mathematical descriptions and models of regularities of all observable objects in the library and information science area. These objects include authors, publications, references, citations, all kinds of texts etc. Bradford's work bases on analyses with journal publications on different subjects in the sciences.

Fundamentally, Bradford law states that literature on any scientific field or subject-specific topic scatters in a typical way. A core or nucleus with the highest concentration of papers - normally situated in a set of few so-called core journals - is followed by zones with loose concentrations of paper frequencies (see Figure 1 for a typical Bradford distribution). The last zone covers the so-called periphery journals which are located in the model far distant from the core subject and normally contribute just one or two topically relevant papers in a defined period. Bradford law as a general law in informetrics can successfully be applied to most scientific disciplines, and especially in multidisciplinary scenarios (Mayr, 2009).

Bradford describes his model in the following:

"The whole range of periodicals thus acts as a family of successive generations of diminishing kinship, each generation being greater in number than the preceding, and each constituent of a generation inversely according to its degree of remoteness." (Bradford, 1934)

Bradford provides in his publications (1934, 1948) just a graphical and verbal explanation of his law. A mathematical formulation has been added later by early informetric researchers. Bradford`s original verbal formulation of his observation has been refined by Brookes (1977) to

$$G(r) = k \ln \left\{ \frac{a+r}{a} \right\} \qquad (1)$$

Where $G(r)$ is the cumulative distribution function, $k$ and $a$ are constants, and $r$ is the rank $1, 2, \ldots n$.

The result of the application of this formula is often called a rank-order distribution of the items in the samples. In the literature we can find different names for this type of distribution, e.g. "long tail distribution", "extremely skewed", "law of the vital few" or "power law" which all show the same properties of a self-similar distribution.

In the past, Bradford law is often applied in bibliometric analyses of databases and collections e.g. as a tool for systematic collection management in library and information science. This has direct influence on later approaches in information science, namely the development of literature databases. The most common known resource which implements Bradford law is the Web of Science (WoS). WoS focuses very strictly on the core of international scientific journals and consequently neglects the majority of publications in successive zones.

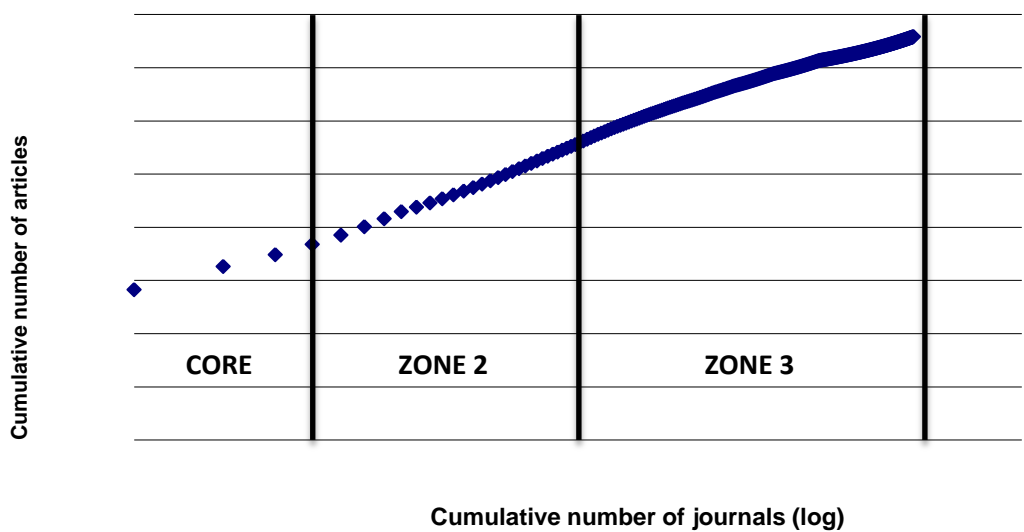

**Figure 1. A typical Bradford distribution: Core, Zone 2 and Zone 3 (so-called periphery). The cumulative number of journals (x-axis) is displayed on a logarithmic scale.**

To conclude this section, Bradford law is relevant for scholarly information systems due to its structuring ability and the possibility to reduce a large document set into a core and succeeding zones. As a consequence, modeling science into a core (producing something like coreness) and a periphery always runs the risk and critic of disregarding important developments outside the core.

*Bradfordizing*
Bradfordizing, originally described by White (1981), is a simple utilization of the Bradford law of scattering model which sorts/re-ranks a result set accordingly to the rank a journal gets in a Bradford distribution. The journals in a search result are ranked by the frequency of their listing in the result set (number of articles in a certain journal). If a search result is bradfordized, articles of core journals are ranked ahead of the journals which contain an average number (Zone 2) or only few articles (Zone 3) on a topic (compare the example in Figure 1). This re-ranking method is interesting because it is a robust and quick way of sorting the central publication sources for any query to the top positions of a result set.

Bradfordizing shows the following advantages: a) a structured view on a result set which is ordered by journals; b) an alternative view on publication sources in an information space which is intuitively closer at the research process than statistical methods (e.g. best match ranking) or traditional methods (e.g. exact match sorting); c) an approach to switch between the search modus e.g. starting with directed term searching and changing to a browsing mode (Bates, 2002) an improvement of relevance distribution between the journal zones, recently investigated (Mayr, 2009).

In principle, the ranking technique Bradfordizing can be applied to any search result with a minimum of 100 documents from one specific document type (e.g. journal articles). Generally Bradfordizing needs 100 or more documents because smaller document sets show too little scattering to divide the result into meaningful zones.

Bates' paper (2002) is interesting in our context because it brings together Bradford's Law (1934), information seeking behavior and IR (compare Wolfram, 2003, Garfield, 1996). Bates postulates "… the key point is that the distribution tells us that information is neither randomly scattered, nor handily concentrated in a single location. Instead, information scatters in a characteristic pattern, a pattern that should have obvious implications for how that information can most successfully and efficiently be sought."

The main task of this paper is to evaluate the effect when applying Bradfordizing to topical document sets from A&I databases. We want to answer the following question: Does Bradfordizing improve the ratio of relevant documents in the first third (core) compared to the second and last third (zone 2 and zone 3, respectively)?

The implementation of Bradfordizing in a typical digital library (DL) should be an alternative ranking option used to re-build and structure a result set. The intention is to list more relevant documents for a topic in the first third of a re-ranked result set. The re-ranking should be interpreted by users as a value-added due to the new structure and the relevance concentration of the listed documents after Bradfordizing. Furthermore Bradfordizing can be a helpful service to positively influence the search process. The opening up of new access paths and possibilities to explore document spaces for academic search questions can be a plausible value-added for users.

In the following section we will describe the research questions and methods used in our study (see Mayr, 2009).

**Methods**

In this paper we seek to answer the following research questions:
1. Is a re-ranking of documents according to Bradfordizing (ranking journal productivity or core journals first) a measurable added value for searchers?

The re-ranking of content to the most frequent sources (extracting the nucleus) can, for example, be a helpful access mechanism for browsing and initial search stages, especially for novice researchers in a discipline. Evaluation of the utility of such a simple re-ranking mechanism is still a desideratum.

2. Are the documents in the nucleus (core journals) of a bradfordized list more often relevant for a topic than items in succeeding zones with lower productivity?

Compared to traditional text-based ranking mechanisms, the bibliometric re-ranking technique Bradfordizing offers a completely new view on result sets, which have not been implemented and tested in heterogeneous database scenarios with multiple collections to date. This requires proving on a larger scale via intellectual assessments.

3. Can Bradfordizing be applied to document sources other than journal articles?

Few analyses show that monograph literature can be successfully bradfordized. But is this a utility for searchers? Other document types (proceedings, grey literature etc.) have to be equally proven.

In our study we focus on document sets from conventional subject-specific A&I databases. We have decided for a laboratory-based IR approach. Intellectual assessments of document relevance were performed following the classical IR evaluation experiments at TREC (e.g. Voorhees, 2007) and Cross-Language Evaluation Forum (CLEF). First of all, the organizers of a retrieval conference like CLEF provide a test collection and a set of topics adequate to this test document corpus. Afterwards, participants apply their individual retrieval algorithms and systems while retrieving these topics (25 different topics each year in CLEF) in the test collection. Each participating retrieval system produces one or more ranked lists (called run) and sends these results back to the organizers. The organizers pool the documents from the retrieval runs for each topic and give the merged document pools away for objective intellectual relevance assessment. All documents in the document pools undergo binary assessment (relevant or irrelevant for a topic) by trained jurors (normally relevance is not binary (see Saracevic, 1975, Mizzaro, 1997 or White, 2007). The jurors perform the assessments on the basis of a short guideline.

We can hypothesize for our experiment: If the ratio of relevant documents, measured in precision (p), is the same in all three equally sized zones, then Bradfordizing has no effect on the distribution of relevant documents in the whole document pool. If the relevance ratio p in the first zone after re-ranking (core) is lower than p in the succeeding zones (zone 2 and zone 3), then Bradfordizing produced a falloff in precision. But if the ratio p of relevant documents in the core is higher than in other zones, and that is what we expect, then Bradfordizing improves the search result (measured in p) and consequently has a positive effect on search.

For this study, topics, documents and intellectual assessments from two evaluation initiatives have been analyzed: document pools from the GIRT-corpus in CLEF and the KoMoHe evaluation project (see Mayr & Petras, 2008). Our study analyzed scientific literature (journal articles and monographs) from social and political sciences, economics, psychology and medical science databases (see Table 1). Documents from the following database were included: SOLIS, SoLit, USB Köln Opac, World Affairs Online, Psyndex and Medline.

**Table 1. Overview of the analyzed topics and documents in the IR experiments.**

|  | **CLEF** | **KoMoHe** |
|---|---|---|
| Project period | 2003-2007 | 2007 |
| Number of topics | 125 | 39 |
| Domain, discipline | Social and political sciences | Social sciences, political sciences, economics, psychology and medical science |
| Assessed documents total | 65,297 | 31,155 |
| Journal articles bradfordized | 18,112 | 17,432 |
| Monographs bradfordized | 11,045 | 4,900 |
| Databases | 2 (1) | 6 |

We retrieved, analyzed and intellectually assessed 164 different standardized topics which yielded more than 96,000 documents from all the above domains. More than 51,000 assessed documents could be bradfordized.

The analysis of the data sets can be divided into three steps.
1. The document types journal articles and monographs are extracted from the document pool. Each document type and topic is analysed separately.
2. Each document set for a topic will be re-ranked according to Bradfordizing and divided into equally sized zones (core, z2 and z3).
3. The relevance assessments of the documents in the three zones are matched and aggregated zone by zone.

Average precision for each topic and zone can be calculated afterwards. We define the precision as the ratio of relevant documents out of all documents.

We calculate the average precision for each zone (core, zone 2 and zone 3) and baseline precision for the whole document pool (see Table 2 for an example).

**Table 2. Example of the applied precision calculation for the CLEF-topic no. 171 "Computers in everyday life".**

|        | Retrieved | Relevant | Precision        |
|--------|-----------|----------|------------------|
| Core   | 73        | 41       | 0.56 (P core)    |
| Zone 2 | 65        | 25       | 0.38 (P z2)      |
| Zone 3 | 70        | 14       | 0.20 (P z3)      |
| Total  | 208       | 80       | 0.38 (P baseline)|

**Results**

The average precision for 164 tested topics from the projects CLEF and KoMoHe increases significantly after Bradfordizing (compare Table 3-6). So we can clearly verify research question 1. In this paper we show only precision values from analyses with journal articles. The largest precision benefit in both datasets is achieved between core and the last zone (zone 3). The improvements in Tables 4 and 6 marked with (*) are statistically significant based on the Wilcoxon signed-rank test and the paired T-Test. The improvements in the KoMoHe tests (see Tables 5, 6) are less significant, but average precision in the core outperforms precision in zone 3 impressively in all test series. Following this result we can clearly verify research question 2.

**Table 3. Average precision for journal articles after re-ranking for five CLEF periods (N=125 topics). Core, Zone 2 (Z2), Zone 3 (Z3) and baseline.**

| CLEF articles | Topics | P core | P Z2  | P Z3  | P baseline |
|---------------|--------|--------|-------|-------|------------|
| 2003          | 25     | 0.294  | 0.218 | 0.157 | 0.221      |
| 2004          | 25     | 0.226  | 0.185 | 0.134 | 0.179      |
| 2005          | 25     | 0.310  | 0.240 | 0.174 | 0.239      |
| 2006          | 25     | 0.288  | 0.267 | 0.244 | 0.265      |
| 2007          | 25     | 0.278  | 0.256 | 0.217 | 0.248      |

**Table 4. Average precision improvements for journal articles for five CLEF periods (N=125 topics). Core, Zone 2 (Z2), Zone 3 (Z3) and baseline.**

| CLEF articles       | P@Core against P@Z3 in % | P@Core against P@Z2 in % | P@Z2 against P@Z3 in % | P@core against baseline in % |
|---------------------|--------------------------|--------------------------|------------------------|------------------------------|
| 2003                | 86.56 (*)                | 34.57 (*)                | 38.63 (*)              | 32.65 (*)                    |
| 2004                | 69.23 (*)                | 22.45                    | 38.20                  | 26.25 (*)                    |
| 2005                | 78.03 (*)                | 29.05 (*)                | 37.95 (*)              | 29.52 (*)                    |
| 2006                | 17.63                    | 7.66                     | 9.27                   | 8.46                         |
| 2007                | 28.18 (*)                | 8.31                     | 18.35                  | 11.77                        |
| Average 2003-2007   | 55.93 (*)                | 20.41 (*)                | 28.48 (*)              | 21.73 (*)                    |

**Table 5. Average precision for journal articles after re-ranking for three KoMoHe tests (N=39 topics). Core, Zone 2 (Z2), Zone 3 (Z3) and baseline.**

| KoMoHe articles | Topics | P core | P Z2 | P Z3 | P baseline |
|---|---|---|---|---|---|
| Test1 | 15 | 0.292 | 0.261 | 0.245 | 0.265 |
| Test2 | 12 | 0.215 | 0.202 | 0.192 | 0.202 |
| Test3 | 12 | 0.700 | 0.644 | 0.587 | 0.642 |

**Table 6. Average precision improvements for journal articles for three KoMoHe tests (N=39 topics). Core, Zone 2 (Z2), Zone 3 (Z3) and baseline.**

| KoMoHe articles | P@Core against P@Z3 in % | P@Core against P@Z2 in % | P@Z2 against P@Z3 in % | P@Core against baseline in % |
|---|---|---|---|---|
| Test1 | 18.82 | 11.75 | 6.32 | 9.84 |
| Test2 | 11.58 | 6.16 | 5.11 | 6.12 |
| Test3 | 19.32 (*) | 8.67 (*) | 9.80 (*) | 9.00 (*) |
| Average Test1-3 | 16.57 (*) | 8.86 | 7.08 (*) | 8.32 (*) |

In general, the precision analyses with monographs in our tests show very similar results. The precision improvements after Bradfordizing (Bradfordizing of publishers) between zones are also positive but less significant than improvements with the journal articles (see research question 3).

**Implementation**

The proposed re-ranking service addresses the problem of oversized result sets by using the bibliometric method Bradfordizing. Bradfordizing re-ranks a result set of journal articles according to the frequency of journals in the result set such that articles of core journals are ranked ahead (see example in Figure 2). This re-ranking method is interesting for retrieval systems because it is a robust and quick way of sorting the central publication sources for any query to the top positions of a result set.

**Figure 2. A bradfordized search for the search term "luhmann". ISSN numbers of journals and their productivity (article counts) are displayed on the left side of the screen. See research prototype under http://multiweb.gesis.org/irsa/IRMPrototype**

The Bradfordizing procedure is implemented in the IRM prototype as a Solr plugin (see Figure 2 and a description of the prototype in Mayr et al., 2011). In a first step the search results are filtered with their ISSN numbers. The next step aggregates all results with an ISSN number. For this step we use a build-in functionality of our prototype engine Solr, the Solr faceting mechanism. Facets in Solr can be defined on any metadata field, in our case the "source" field of our databases. The journal with the highest ISSN count gets the top position in the result. The second journal gets the next position, and so on (see example in Figure 2). This procedure is an exact implementation of the original Bradfordizing approach. In the last step, the document ranking step, our current implementation works with a simple boosting mechanism. The frequency counts of the journals are used as boosting factors for documents in these journals. The numerical ranking value from the original tf-idf ranking of each document is multiplied with the frequency count of the journal (see Schaer, 2011). The result of this multiplication will be taken as ranking value for the final document ranking.

In principle, this ranking technique can be applied to any search result providing qualitative metadata (e.g. journal articles in literature databases). Generally, Bradfordizing needs 100 or more documents because smaller document sets often show too little scattering to divide the result into meaningful zones. Bradfordizing can be applied to document types other than journal article, e.g. monographs (cf. Worthen, 1975; Mayr, 2008, 2009). Monographs e.g. provide ISBN numbers which are also good identifiers for the Bradfordizing analysis.

To conclude, our implementation of re-ranking by Bradfordizing is a simple approach which is generic, adaptable to various document types and quickly implementable with build-in functionality. The only precondition for the application is the existence of qualitative metadata to assure precise identification and access to the documents. An evaluation of the value-added services of Bradfordizing and other approaches has been published recently by Mutschke et al. (2011).

**Discussion**
The discussion of the re-ranking method Bradfordizing will focus on possible added-values and the positive and negative effects of this method. Some added-values appear very clearly. On an abstract level, re-ranking by Bradfordizing can be used as a compensation mechanism for enlarged search spaces with interdisciplinary document sets. Bradfordizing can be used in favor of its structuring and filtering facility. Our analyses show that the hierarchy of the result set after Bradfordizing is a completely different one compared to the original ranking. The user gets a new result cutout with other relevant documents which are not listed in the first section (in our experiment the top 10 documents) of the original list. Furthermore, Bradfordizing can be a helpful information service to positively influence the search process, especially for searchers who are new on a research topic and don't know the main publication sources in a research field. The opening up of new access paths and possibilities to explore document spaces can be a very valuable facility. Additionally, re-ranking via bradfordized documents sets offer an opportunity to switch between term-based search and the search mode browsing. It is clear that the approach will be provided as an alternative ranking option, as one additional way or stratagem to access topical documents (cf. Bates, 2002).

Interesting in this context is a statement by Bradford where he explains the utility of the typical three zones. The core and zone 2 journals are in his words "obviously and a priori relevant to the subjects", whereas the last zone (zone 3) is a very "mixed" zone, with some relevant journals, but also journals of "very general scope" (Bradford, 1934). Pontigo and Lancaster (1986) come to a slightly different conclusion of their qualitative study. They investigated that experts on a topic always find a certain significant amount of relevant items in the last zone. This is in agreement with quantitative analyses of relevance assessments in the Bradford zones (Mayr, 2009). The study shows that the last zone covers significantly less often relevant documents than the core or zone 2. The highest precision can very constantly be found in the core.

To conclude, modeling science into a core and a periphery – the Bradford approach – always runs the risk and critic of disregarding important developments outside the core. Hjorland and Nicolaisen (2005) recently started a first exploration of possible side effects and biases of the Bradford methods. They criticized that Bradfordizing favours majority views and mainstream journals and ignores minority standpoints. This is a serious argument, because by definition, journals which publish few papers on specific topics have very little chance to get into the core of a more general topic. A counter-argument could be that the Bradfordizing approach is just an application which is working on existing document sets. The real problem is situated before, in the development of a data set, especially in the policy of a database producer.

**Conclusions**
An evaluation of the method and its effects was carried out in two laboratory-based information retrieval experiments (CLEF and KoMoHe) using a controlled document corpus and human relevance assessments (see Ingwersen & Järvelin, 2005 for pros and cons of this methodology). The results show that Bradfordizing is a very robust and promising method for re-ranking the main document types (journal articles and monographs) in today's digital libraries (DL). The IR tests show that relevance distributions after re-ranking improve at a significant level if articles in the core are compared with articles in the succeeding zones. The items in the core are significantly more often assessed as relevant, than are items in zone 2 or zone 3. The largest increase in precision can typically be observed between core and zone 3. This has been called the Bradfordizing effect.

The results of our study can also be seen as a coalescence of Bradford Law in so far as Bradford did not postulate or observe a relevance advantage in the core. In Bradford's eyes all documents in his bibliographies were "relevant to a subject". His focus was the scattering of documents across journals, not the relevance distribution between document zones. According to Saracevic (1975), Bradford (1934) was one of the first persons to use the term relevant in our context ("relevant to a subject"). The results in this study show that articles in core journals are valued more often relevant than articles in succeeding zones (compare Garfield, 1996). This is an extension to the original conception of relevance distribution in the zones by Bradford. As we can empirically see, bibliometric distributions like Bradford distributions can also be described as "relevance related distributions" (Saracevic, 1975). The examination of relevance concentrations in our test series (CLEF and KoMoHe) show that there is not a massive concentration of relevant articles in the core, rather it is more a continuously decreasing of average precision from core to zone 3.

The relevance advantage in the core can probably be explained in that a) core journals publish more state-of-the-art articles, b) core journals are more often reviewed by peers in a certain field and c) core journals cover more aspects of the searched topic than journals in the peripheral zones. Further research is needed to clarify these questions.

**Further research**

After evaluating the positive relevance effect of Bradfordizing, our next goal is to go automatically from directed searching into a browsing mode. Starting with a subject-specific descriptor search, we will treat the query with our heterogeneity modules (Mayr & Petras, 2008) to transfer descriptor terms into a multi-database scenario. In a second step, the result lists from the distributed databases are combined, merged and re-ranked by users e.g. according to Bradfordizing. Step 3 could be the extraction of a result set of documents in the Bradford nucleus which can be delivered for browsing or other search stratagems. This browsing modus, based on automatically bradfordized lists, can be compared to the search technique which Bates terms "journal run."

The exploration of possible side effects and bias (see e.g. Nicolaisen & Hjorland, 2007) of this promising re-ranking method will be a next step. Recently Nicolaisen & Hjorland have criticized Bradfordizing: "Bradford analyses function discriminatorily against minority views … Bradford analysis can no longer be regarded as an objective and neutral method." This has to be proven on a larger empirical basis.

A comparison with other ranking and re-ranking methods would be highly desired. Techniques like bibliometric re-ranking (e.g. Bradfordizing described in this paper) or the application of social-network analysis techniques (e.g. co-authorship relationships in Mutschke, 2003) or other combinations of value-added services can and should be applied in digital libraries (DL) to improve IR (White 2005, 2007). Further research will focus on the implementation and evaluation of the method in a live system with different modules for improving retrieval (see Mutschke et al, 2011).

**Acknowledgement**
The KoMoHe project at GESIS ("Competence Center Modeling and Treatment of Semantic Heterogeneity") was the starting point and background of this study. The KoMoHe project was funded by the German Federal Ministry for Education and Research (BMBF) grant number 523-40001-01C5953. The retrieval prototype has been developed in the DFG project "Value-Added Services for Information Retrieval" (IRM) under project number: INST 658/6-1.